\documentclass[11pt]{article}
\pdfoutput=1

\usepackage{lmodern,indentfirst}
\usepackage{amsmath,amssymb,amsthm,amscd}
\usepackage{graphicx}
\usepackage{subfig}
\usepackage[svgnames]{xcolor}
\usepackage{slashed}
\usepackage{dsfont}
\usepackage{bbold}
\usepackage{multirow}
\usepackage{booktabs}
\usepackage{color}
\usepackage[body={6.5in, 9in}, right=1in, top=1in]{geometry}
\usepackage{cite}
\usepackage[colorlinks=true,linkcolor=Blue,citecolor=Green,urlcolor=violet]{hyperref}

\usepackage{soul}

\newcommand{\bs}[1]{\boldsymbol{#1}}
\def\({\left(}
\def\){\right)}
\def\[{\left[}
\def\]{\right]}
\def\<{\left\langle}
\def\>{\right\rangle}
\def\rd{{\rm d}}
\newcommand{\dx}[2]{{\rm d}^{#1} #2}

\usepackage{soul}

\newcommand{\figref}[1]{Fig.~\ref{#1}}

\begin{document}
\def\thefootnote{\fnsymbol{footnote}}

\begin{center}
\LARGE{\textbf{On the IR-Resummation in the EFTofLSS}} \\ 
\vspace{1cm}

\large{Leonardo Senatore$^{\rm 1, 2}$ and Gabriele Trevisan$^{\rm 3}$}\\

\vspace{.5cm}

\small{
\textit{$^{1}$ Stanford Institute for Theoretical Physics, Stanford University,\\ Stanford, CA 94306}\\
\textit{$^{2}$ Kavli Institute for Particle Astrophysics and Cosmology, Physics Department and SLAC,\\ Menlo Park, CA 94025}\\
\textit{$^{3}$ Center for Cosmology and Particle Physics, Department of Physics, New York University,\\ New York, NY 10003, USA}
}
\end{center}

\vspace{.5cm}
 \begin{center}
\large \textbf{Abstract}
 \end{center}
 \vspace{.5cm}
We propose a simplification for the IR-resummation scheme of~\cite{Senatore:2014via} and also include its next-to-leading order corrections coming from the tree-level three-point function of the long displacement field. First we show that the new simplified formula shares the same properties of the resummation of~\cite{Senatore:2014via}. In Fourier space, the IR-resummed power spectrum has no residual wiggles and the two-loop calculation matches the non-linear power spectrum of the Dark Sky simulation at $z=0$ up to $k\simeq0.34\,h\,\text{Mpc}^{-1}$ within cosmic variance. Then, we find that the additional subleading terms (although parametrically infrared-enhanced) modify the leading-order IR-resummed correlation function only in a marginal way, implying that the IR-resummation scheme can robustly predict the shape of the BAO peak.

\def\thefootnote{\arabic{footnote}}
\setcounter{footnote}{0}
\newpage

\section{Introduction}

The recent years have seen a rapid implementation of Effective Field Theory (EFT) methods to the study of the Large Scale Structure (LSS) \cite{Baumann:2010tm,Carrasco:2012cv,Pajer:2013jj,Mercolli:2013bsa,Senatore:2014vja,Porto:2013qua,Senatore:2014eva,Lewandowski:2015ziq,Baldauf:2015aha,Assassi:2015jqa,Baldauf:2014qfa,Angulo:2014tfa,Perko:2016puo,Foreman:2015lca}. The EFT formalism cures various UV-sensitive perturbative calculations of Eulerian and Lagrangian Perturbation Theory (SPT and LPT, see \cite{Bernardeau:2001qr} for a review), including the Zel'dovich approximation \cite{Zeldovich:1969sb}. The corrections introduced by the EFT are already relevant in the one-loop calculation and fix the uncontrollable contribution from UV-physics into loop integrals (which is not amenable to a perturbative calculation). 

The EFT of LSS was initially developed in the Eulerian framework (and later also in Lagrangian space) and has been used to calculate the matter power spectrum up to two loops \cite{Baumann:2010tm,Carrasco:2012cv,Pajer:2013jj,Mercolli:2013bsa,Senatore:2014vja,Carrasco:2013sva,Carrasco:2013mua,Baldauf:2015aha} and the bispectrum up to one loop \cite{Baldauf:2014qfa,Angulo:2014tfa}. In particular \cite{Baldauf:2014qfa,Foreman:2015lca,Cataneo:2016suz} have shown that the EFTofLSS ameliorates the reach in momentum $k$ of the theory compared to simulations.

\begin{figure}[!htb]
\begin{center}
\includegraphics[width=0.49 \textwidth]{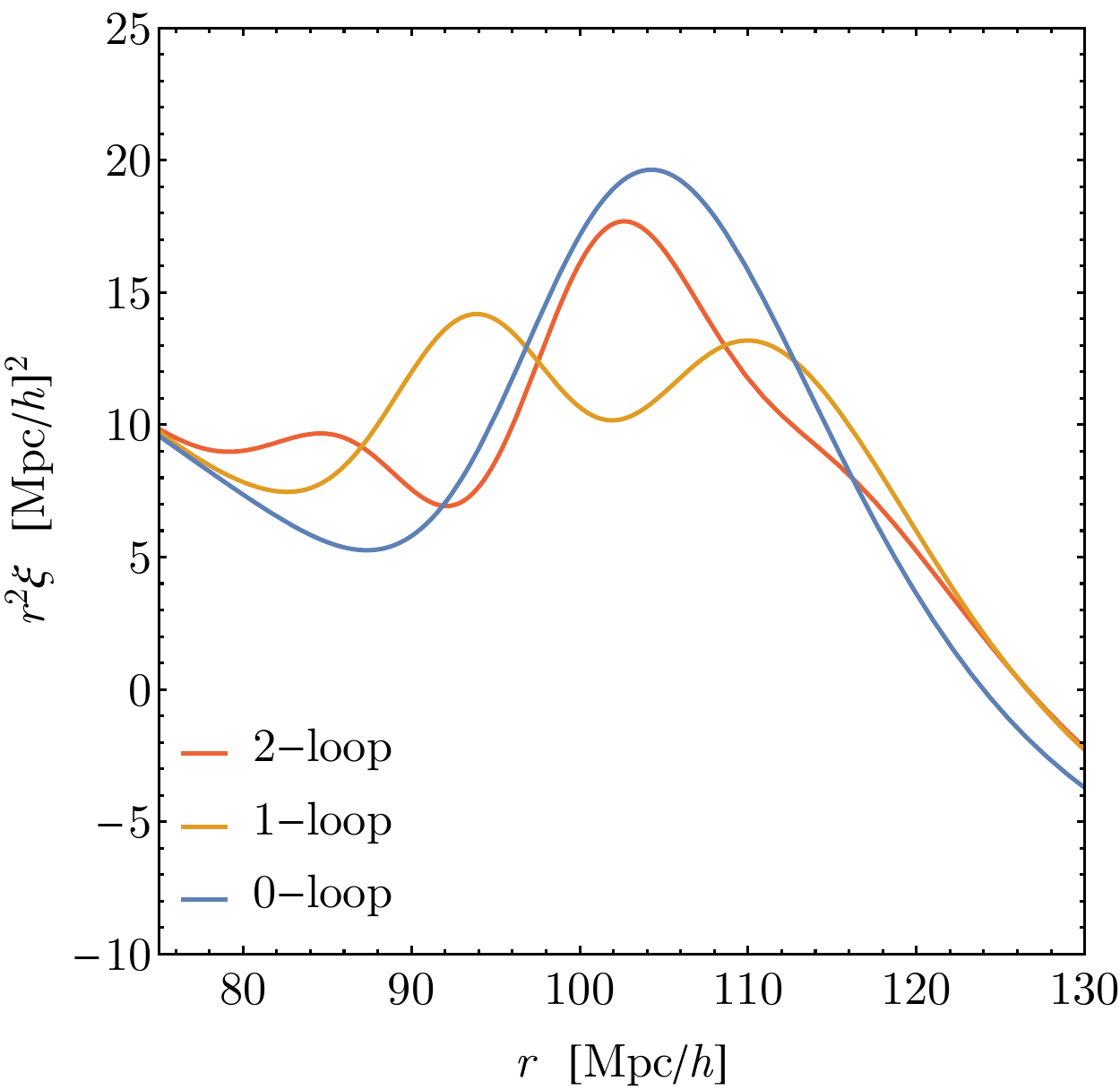}
\end{center}
\caption{\small {\em Predictions of Eulerian PT for the correlation function around the BAO peak at zero-, one- and two-loop without IR-Resummation.}}
\label{fig.PTBAO}
\end{figure}

However, regardless of the improvement in Fourier space, it is well known that the prediction of Eulerian PT in real space around the BAO peak, $\ell_\text{BAO}\simeq 100\,h^{-1}\text{Mpc}$, is rather unsatisfactory beyond the linear level. Looking for example at \figref{fig.PTBAO}, one can easily see that the one-loop and two-loop calculations fail to improve the prediction of the shape and position of the BAO peak, even though for our universe the relevant modes are in the perturbative regime. As already explained in Refs.~\cite{Senatore:2014via, Baldauf:2015xfa, Tassev:2013rta}, the reason for this failure can be traced back to the expansion parameters controlling Eulerian PT. In the Eulerian treatment, one expands in powers of the linear fluctuations. In practice, this amounts to expanding in different parameters; to be specific, let us take the one-loop contribution which goes as
\begin{equation}\label{eq.P1loop}
\begin{split}
P_{1-\text{loop}}(k)&=P_{22}(k)+P_{13}(k)\\
&=\int \frac{\rd^3p}{(2\pi)^3} \[2F_2^2(\bs{p},\bs{k}-\bs{p})P_{\text{lin}}(|\bs{k}-\bs{p}|)+6 F_3(\bs{p},-\bs{p},\bs{k})P_{\text{lin}}(|\bs{k}|)\]P_{\text{lin}}(p),
\end{split}
\end{equation} 
where the $F_n$ are the symmetrized kernels of SPT \cite{Bernardeau:2001qr}. 
It is instructive to see how a mode $\bs{p}$ in the loop integral affects the correction to the linear power spectrum. Let us start from the UV. From Eq.~(\ref{eq.P1loop}), the dominant contribution to $P_{1-\text{loop}}$ as $p\rightarrow\infty$ comes from $P_{13}$ and reads
\begin{equation}\label{eq.epsilonpsi>}
\begin{split}
P_{13}(k)&\sim  P_{\text{lin}}(k) \, k^2\int_{p\gg k} \frac{\rd^3p}{(2\pi)^3} \frac{P_{\text{lin}}(p)}{p^2}\\
&\equiv P_{\text{lin}}(k)\,\epsilon_{s_>}.
\end{split}
\end{equation}
On the other hand, in the IR, the contribution of a mode $\bs{p}$ goes as
\begin{equation}\label{eq.epsilondelta<}
\begin{split}
P_{22}(k)+P_{13}(k)&\sim P_{\text{lin}}(k)\,\int_{p\ll k} \frac{\rd^3p}{(2\pi)^3} P_{\text{lin}}(p)\\
&\equiv P_{\text{lin}}(k)\, \epsilon_{\delta_<}.
\end{split}
\end{equation} 
As one can see, the corrections to the power spectrum are actually controlled by more than one parameter: in the UV these depends on the variance of short displacement $s_>(p)\sim\delta(p)/p$, while in the IR these depends on the variance of long density fluctuation $\delta_< (p)$. 
It is not surprising that very long displacement-modes $s_<$ do not appear in the equal-time two-point function, since displacements larger than the correlation length are unobservable because of the equivalence principle \cite{Scoccimarro:1995if,Jain:1995kx,Creminelli:2013mca,Carrasco:2013sva}.  
However, what is more interesting is that in SPT the absence of these modes comes through a cancellation between different diagrams (e.g. at one loop between $P_{22}$ and $P_{13}$). Explicitly, going back to Eq.~(\ref{eq.P1loop}) and taking the IR limit inside the integral, at one-loop one gets
\begin{equation}\label{eq.P1loopIR}
\begin{split}
P_{1-\text{loop}}(k)&\sim\frac{1}{2}\int_{p\ll \Lambda} \frac{\rd^3p}{(2\pi)^3} \frac{\(\bs{p}\cdot\bs{k}\)^2}{p^4}\[P_{\text{lin}}(|\bs{k}-\bs{p}|)+P_{\text{lin}}(|\bs{k}+\bs{p}|)-2P_{\text{lin}}(|\bs{k}|)\]P_{\text{lin}}(p).
\end{split}
\end{equation}
But which cut-off $\Lambda$ should we consider? For a featureless linear power spectrum, say $P_{\text{lin}}(k)\propto k^n$, one can easily see that for $p\ll\Lambda\simeq k$,
\begin{equation}\label{eq.?}
\begin{split}
\[P_{\text{lin}}(|\bs{k}-\bs{p}|)+P_{\text{lin}}(|\bs{k}+\bs{p}|)-2P_{\text{lin}}(|\bs{k}|)\]\sim P_{\text{lin}}(k) \frac{p^2}{k^2},
\end{split}
\end{equation}
matching the limit in Eq.~(\ref{eq.epsilondelta<}). This squares with the naive expectations that modes of the displacement in the IR with respect to $k$ cannot affect equal time correlators. 

However our Universe is not featureless because of the BAO (an oscillation of frequency $2\pi\ell_{\text{BAO}}^{-1}$ in the power spectrum) which translates to a bump in the correlation function at a separation $r\simeq\ell_{\text{BAO}}$. In the presence of such a feature the square bracket in Eq.~(\ref{eq.P1loopIR}) now reads
\begin{equation}\label{eq.?}
\begin{split}
\[P^w_{\text{lin}}(|\bs{k}-\bs{p}|)+P^w_{\text{lin}}(|\bs{k}+\bs{p}|)-2P^w_{\text{lin}}(|\bs{k}|)\]&=2\int\rd^3 r\, \xi^w(r) e^{-i \bs{k}\cdot \bs{r}} [\cos\(\bs{p}\cdot\bs{r}\)-1]\\
&\sim P^w_{\text{lin}}(k) \(\cos\(p\,\ell_{\text{BAO}}\)-1\),
\end{split}
\end{equation}
so that the cancellation of long displacement modes happens only for modes $p$ in the IR of $\ell_{\text{BAO}}^{-1}$, that is $p\ll\Lambda\simeq\ell_{\text{BAO}}^{-1}$, and \emph{not} of $k$. Furthermore, the correction to the linear wiggly component $P^w_{\text{lin}}$ is parametrically different from the previous correction, Eqs.~(\ref{eq.epsilondelta<}), since it receives a contribution from modes $\ell_{\text{BAO}}^{-1}\lesssim p \lesssim k$ that goes as
\begin{equation}\label{eq.P1loopW}
\begin{split}
P^w_{1-\text{loop}}(k)&
\sim  P^w_{\text{lin}}(k)\,k^2 \int_{\ell_{\text{BAO}}^{-1}\lesssim p \lesssim k} \frac{\rd^3p}{(2\pi)^3} \frac{P_{\text{lin}}(p)}{p^2}\\
&\equiv P^w_{\text{lin}}(k) \epsilon_{s_<}.
\end{split}
\end{equation}
For our universe this parametrically \emph{IR-enhanced} contribution is large since $\epsilon_{s_<}\simeq 1$ for modes in the support of the wiggly power spectrum ($0.05\lesssim k^w\lesssim 0.3$). This is why the convergence of Eulerian Perturbation theory is rather slow, as one can see in Fig.~\ref{fig.PTBAO}.\footnote{Notice however that the dependence on long displacement field is analytic, so PT \emph{will converge} to the correct answer after enough insertions of  $\epsilon_{s_<}$ (this is not the case for insertions of $\epsilon_\delta$, since the dependence of the full result is non-analytic, so PT \emph{will not converge}, though, for $k\ll k_\text{NL}$, the approximation of the EFT to the correct result will be very accurate).}  

To resolve this issue, taking inspiration from Lagrangian Perturbation Theory, which does not expand in the displacement field and therefore performs remarkably well around the BAO peak \cite{Carlson:2012bu}, Ref.~\cite{Senatore:2014via} proposed a technique to resum the IR-enhanced contributions of the perturbative calculation within the Eulerian picture. The goal of this paper\footnote{We have made the code used for the plots in this paper available on the \href{http://stanford.edu/~senatore/}{EFTofLSS repository}.} is to improve upon the work of \cite{Senatore:2014via} by including the next-to-leading (NLO) corrections to the IR-resummation, which accounts for the non-linear evolution of the long displacement field and the mode-coupling between long and short modes. 

Other techniques have also been developed. For example Refs.~\cite{Vlah:2015sea,Vlah:2014nta} have studied loop corrections in LPT. Refs. \cite{Blas:2015qsi,Blas:2016sfa} have developed a different perturbative approach (equivalent order by order to Eulerian perturbation theory) within which IR-enhanced contribution have been identified and resummed. 

\section{IR-resummation}\label{sec.lagrangian}
\subsection{Preliminaries}
To better understand the origin of the IR-resummation of~\cite{Senatore:2014via}, it is better to start from Lagrangian space. In the Lagrangian picture, the clustering of dark matter particles is determined by the displacement field $\bs{s}$ of each particle from its initial position $\bs{q}$. The position of a particle in real space, at time $t$, is given by
\begin{equation}\label{eq.ELMap}
\begin{split}
\bs{x}(\bs{q},t)=\bs{q}+\bs{s}(\bs{q},t).
\end{split}
\end{equation}
Assuming a uniform initial density field, and that $\bs{q}$ is a continuous variable, due to conservation of mass
\begin{equation}
\begin{split}
\bar\rho\,\rd^3q=\rho(\bs{x},t)\,\rd^3x,
\end{split}
\end{equation}
one can relate the overdensity in real space to the displacement via the change of coordinates in Eq.~(\ref{eq.ELMap}),
\begin{equation}
\begin{split}
1+\delta(\bs{x},t)=\int\rd^3q\,\delta_D^3(\bs{x}-\bs{q}-\bs{s}(\bs{q},t)),
\end{split}
\end{equation}
where $\delta_D$ is the Dirac $\delta$-function. Going to Fourier space, for $k\neq0$, leads to 
\begin{equation}
\begin{split}
\delta(\bs{k},t)=\int\rd^3q\,e^{-i\bs{k}\cdot \bs{q}-i\bs{k}\cdot \bs{s}(\bs{q},t)}.
\end{split}
\end{equation}
With these definitions, the matter power spectrum reads
\begin{equation}\label{eq.P}
\begin{split}
P(k)=\int \rd^3q\, e^{-i\bs{k}\cdot\bs{q}}\<e^{-i\bs{k}\cdot\bs{\Delta}(\bs{q})}\>-\(2\pi\)^3\delta_D^3(\bs{k}),
\end{split}
\end{equation}
where $\bs{q}=\bs{q}_1-\bs{q}_2$ is the separation between between two particles, and $\bs{\Delta}(\bs{q})=\bs{s}(\bs{q}_1)-\bs{s}(\bs{q}_2)$.\\
In the Lagrangian EFT~\cite{Porto:2013qua}, the effect of non-linear short-wavelength mode is encoded in the fact that the particles have a finite size (of oder of the non-linear scale). However, as in~\cite{Senatore:2014via}, since here we are interested in the resummation of the long-wavelength displacements, we can neglect this subtlety, as it is associated to the UV physics.

Let us explore the structure of the perturbative expansion. The expression in Eq.~(\ref{eq.P}) involves the expectation value of an exponential, which can be to evaluate with the cumulant expansion theorem as follows 
\begin{equation}\label{eq.K}
\begin{split}
\< e^{-i\bs{k}\cdot\bs{\Delta}(\bs{q})}\>
 =\exp\[\sum^\infty_{n=1}\frac{(-i)^n}{n!}\<(\bs{k}\cdot\bs{\Delta}(\bs{q}))^n\>_c\]=K(\bs{k},\bs{q}),
\end{split}
\end{equation}
where $\<X^n\>_c$ is the $n$th-order cumulant of the stochastic variable $X$, usually calculated as a perturbative expansion of connected diagrams. To calculate the two-point function of the density, one in principle needs all the correlation function of the displacement. Schematically one has to calculate terms like
\begin{equation}
\begin{split}
\<s(q_1)s(q_2)\>, \quad \<s(q_1)s(q_2)s(q_3)\>, \quad\dots \quad .
\end{split}
\end{equation}
From the continuity equation at linear order one has
\begin{equation}\label{eq.?}
\begin{split}
\bs{s}(p)\simeq \bs{s}_\text{lin}(p)= i\frac{\bs{p}}{p^2}\delta_\text{lin} (p),
\end{split}
\end{equation}
and only the two-point function is non-zero, so that
\begin{equation}\label{eq.?}
\begin{split}
K(\bs{k},\bs{q})\simeq \exp\[\frac{1}{2}\int \frac{\rd^3p}{(2\pi)^3} \frac{\(\bs{p}\cdot\bs{k}\)^2}{p^4} P_\text{lin}(p)\(e^{i \bs{p}\cdot\bs{q}}-1\)\],
\end{split}
\end{equation}
which once expanded contains as term in $k^2/p^2$ as in Eq.~(\ref{eq.P1loopW}). 

At higher order in LPT, the $n$-th order contribution to the displacement goes as 
\begin{equation}\label{eq.pertdispl}
\begin{split}
\bs{s}^{(n)}(p)\sim i\frac{\bs{p}}{p^2}\,\delta^{n}_\text{lin} (p)\sim \bs{s}_\text{lin} (p)\,\delta^{n-1}_\text{lin} (p).
\end{split}
\end{equation}
In our paper, we are interested in resumming the effect of the long displacement, which we call ``IR-enhanced". The expressions in Eq.~(\ref{eq.K}) and (\ref{eq.pertdispl}) therefore tell us that, beyond the leading term $\epsilon_{s_<}$, there is a subleading IR-enhanced term which involves non-linearities in $\delta$.
With this counting, the next-to-leading order IR-enhanced correction to the power spectrum of modes of order $k_\text{short}$ comes from the tree-level three-point function of the long displacements. Let us take for simplicity a scaling universe with $P(k)\propto \(k/k_{\text{NL}}\)^n$. For our universe, $-1<n<-2$. this contribution compared to the leading-order term scales as 
\begin{equation}\label{eq.?}
\begin{split}
\frac{k_\text{short}^3 \langle s_< s_< s_<\rangle}{k_\text{short}^2 \langle s_< s_< \rangle}\sim\frac{\frac{k_\text{short}}{k_\text{long}}\epsilon_{s_<}\epsilon_{\delta_{<,\text{low}}}}{\epsilon_{s_<}}= \frac{k_\text{short}}{k_\text{long}}\epsilon_{\delta_{<,\text{low}}}\sim \frac{k_\text{short}}{k_\text{NL}} \(\frac{k_\text{long}}{k_\text{NL}}\)^{n+2}\ll 1,
\end{split}
\end{equation}
which is small, but it is IR-enhanced with respect to the infrared part of the loop-correction to the two-point function of $\delta$ as 
\begin{equation}\label{eq.?}
\begin{split}
\frac{k_\text{short}^3 \langle s_< s_< s_<\rangle}{\epsilon_{\delta_{<,\text{low}}}}\sim\frac{\frac{k_\text{short}}{k_\text{long}}\epsilon_{s_<}\epsilon_{\delta_{<,\text{low}}}}{\epsilon_{\delta_{<,\text{low}}}}=\frac{k_\text{short}}{k_\text{long}}\epsilon_{s_<}.
\end{split}
\end{equation}
However one should also worry about the contribution of density non-linearities due to short-wavelength modes. This ratio goes as
\begin{equation}\label{eq.NLOoverLoopShort}
\begin{split}
\frac{k_\text{short}^3 \langle s_< s_< s_<\rangle}{\epsilon_{\delta_{<,\text{high}}}}\sim\frac{\frac{k_\text{short}}{k_\text{long}}\epsilon_{s_<}\epsilon_{\delta_{<,\text{low}}}}{\epsilon_{\delta_{<,\text{high}}}}\sim \(\frac{k_\text{long}}{k_\text{short}}\)^{n+2}\epsilon_{s_<},
\end{split}
\end{equation}
which in principle would be large if  $n\lesssim-2$. In our universe $\epsilon_{s_<}\approx1$ and the BAO wiggles are in a region of $k$ where  $-1\lesssim n \lesssim-2$, so that the NLO terms are IR-enhanced with respect to the long-wavelength loop corrections, but are smaller than the short-wavelength loop correction involving $\epsilon_{\delta_{<,\text{high}}}$. Since corrections in $\epsilon_\delta$ are not resummed but instead are treated perturbatively, Eq.~(\ref{eq.NLOoverLoopShort}) shows that it is not quite justified to include in the resummation the effect associated with the three-point function of the long displacements. In fact, in doing so, one would include terms that are not definitely larger than what we neglect. However, since resumming the long NLO terms is quite simple (as we will show later in Eq.~(\ref{eq.K0W})), we will include them and confirm that their effect is quite marginal in Fig.~\ref{fig.corr}.\\

For what follows we will need the two- and three-point functions of the displacement field. By rotational invariance these can be decomposed into their ``spin-0", ``spin-2" and ``spin-1", ``spin-3" components respectively. In formulae
\begin{equation}\label{eq.ABdecomposition}
\begin{split}
A_{ij}(\bs{q})&=A_0(q) \delta_{ij}+A_2(q) \hat q_i \hat q_j\\
B_{ijk}(\bs{q})&=B_1(q) \delta_{\{ij}\hat q_{k\}} +B_3(q) \hat q_i \hat q_j \hat q_k,
\end{split}
\end{equation}
where the functions $A_0$, $A_2$, $B_1$ and $B_3$ can be calculated perturbatively \cite{Matsubara:2007wj}, and are reported in App.~\ref{app.AB}.

\subsection{A simplified resummation procedure}\label{sec.main}
Before turning to the next-to-leading order terms, let us introduce a simplified resummation procedure with respect to \cite{Senatore:2014via}.
We are interested in resumming (i.e. keep exponentiated) only terms which contains long-wavelength displacements and their NLO corrections, and expand in all other terms. As usual, we can then do a two steps procedure: first, since we want to resum IR contributions  we can neglect all UV contributions encapsulated in the EFT terms; second, once the IR-resummation is done, we can reintroduce the EFT terms perturbatively. Following \cite{Senatore:2014via}, we will indicate the term in Eq.~(\ref{eq.K}) when expanded up to order $N$ in $\epsilon_\delta$ as
\begin{equation}
\begin{split}
K(\bs{k},\bs{q})\big|_N\qquad\qquad(\text{resumed}),
\end{split}
\end{equation}
and when expanded up to order $N$ in all parameters (which corresponds to the usual expansion in $P_{\text{lin}}$) as
\begin{equation}
\begin{split}
K(\bs{k},\bs{q})\big|\big|_N\qquad\qquad(\text{not resumed}).
\end{split}
\end{equation}
At leading order, the terms that we wish to keep exponentiated from Eq.~(\ref{eq.K}) are
\begin{equation}\label{eq.K0}
\begin{split}
K_0(\bs{k},\bs{q})\equiv \left.K(\bs{k},\bs{q})\right|_0&=\exp\[-\frac{1}{2}k_ik_j A_{ij}^{\text{IR}}(\bs{q})\].
\end{split}
\end{equation}
where $A^{\text{IR}}$ and $B^{\text{IR}}$ contain only the IR modes\footnote{Explicitly, we dump the integrals in Eq.~(\ref{eq.xi}) by a factor $ \exp\(-p^2/\Lambda^2_\text{IR}\)$, where $\Lambda_\text{IR}\approx 0.1\,h\,\text{Mpc}^{-1}$.} of the tensors $A$ and $B$. \\
Since $K$ and $K|_0$ are equal at zero-order in $\epsilon_\delta$, $K|_0$ already contains all the information we wish to resum. Therefore we can use the following trick
\begin{equation}
\begin{split}
\left.K\right|_N &\simeq K_0\cdot\left.\left.\frac{K}{K_0}\right|\right|_N\\ 
&=K_0 \sum_{j=0}^N \left.\left.K_0^{-1}\right|\right|_{N-j} \cdot K_{j}\\  
&= \sum_{j=0}^N \left.\left.F\right|\right|_{N-j}\cdot K_{j},
\end{split}
\end{equation}
where $K_{j}$ is the $j$-th term in the Taylor series in both parameters $\epsilon_s$ and $\epsilon_\delta$ (that is in power of $P_\text{lin}$), and we omitted the dependence on $\bs{k}$ and $\bs{q}$. For convenience we have also defined
\begin{equation}\label{eq.defF}
\begin{split}
\left.\left.F\right|\right|_{N-j}\equiv K_0 \cdot \left.\left.K^{-1}_0\right|\right|_{N-j}.
\end{split}
\end{equation}
With all this, the two-point function up to order $N$ in $\epsilon_\delta$ reads
\begin{equation}
\begin{split}
P(k)=\int \rd^3 q \,e^{-i \bs{k}\cdot\bs{q}}\;  \sum_{j=0}^N \left.\left.F(\bs{k},\bs{q})\right|\right|_{N-j}\cdot K_{j}(\bs{k},\bs{q})-\(2\pi\)^3\delta_D^3(\bs{k}).
\end{split}
\end{equation}
Up to this point the resummation procedure coincides with the one in \cite{Senatore:2014via}. However here we will adopt a different approach for which it is more convenient to Fourier transform this expression and obtain the two-point function in real space
\begin{equation}\label{eq.master}
\begin{split}
1+\xi(r)=\int\frac{\rd^3 k\,\rd^3 q}{(2\pi)^3} \,e^{i \bs{k}\cdot(\bs{r}-\bs{q})}\;  \sum_{j=0}^N \left.\left.F(\bs{k},\bs{q})\right|\right|_{N-j}\cdot K_{j}(\bs{k},\bs{q}).
\end{split}
\end{equation}
At this point one can notice that it is a consistent approximation to expand $F(\bs{k},\bs{q})$ around $\bs{q}\simeq\bs{r}=\bs{x}_1-\bs{x}_2$, since this involves gradients of the displacement field which are parametrically suppressed with respect to the terms we are resumming (indeed, they scale as $\epsilon_{\delta_<}$). Let us check this explicitly for the two-point correlation function appearing in the exponent of Eq.~(\ref{eq.K0}). Inverting the relation in Eq.~(\ref{eq.ELMap}), one can formally expand $\bs{s}(\bs{q})$ as
\begin{equation}\label{eq.s(x)}
\begin{split}
\bs{s}(\bs{q}_1)=\bs{s}(\bs{x_1}-\bs{s}(\bs{q}_1))\simeq\bs{s}(\bs{x}_1)-\bs{s}(\bs{x}_1)\cdot\nabla\bs{s}(\bs{x}_1),
\end{split}
\end{equation}
 to obtain
\begin{equation}\label{eq.?}
\begin{split}
\<s_i(\bs{q}_1)s_j(\bs{q}_2)\>=\<s_i(\bs{x}_1)s_j(\bs{x}_2)\>-\<s_k(\bs{x}_1) \nabla_k s_i(\bs{x}_1)s_j(\bs{x}_2) \>-\<s_i(\bs{x}_1)s_k(\bs{x}_2)\nabla_k s_j(\bs{x}_2)\>.
\end{split}
\end{equation}
The last two terms scale  as the one-loop correction involving $\epsilon_{\delta_{<,\text{low}}}$ to the two-point function which, as we have already argued, are parametrically suppressed with respect to the third order cumulant that we will keep in at the end of Subsec.~\ref{sec.main}. With all this, Eq.~(\ref{eq.master}) simplifies to
\begin{equation}\label{eq.masterapprox}
\begin{split}
\xi(r)=\int \frac{\rd^3 k}{(2\pi)^3} \;e^{i \bs{k}\cdot\bs{r}}\;  \sum_{j=1}^N \left.\left.F(\bs{k},\bs{r})\right|\right|_{N-j} P^{\text{E}}_{j}(\bs{k}),
\end{split}
\end{equation}
where we used the well known fact that 
\begin{equation}
\begin{split}
\int \dx{3}{q}\;e^{i \bs{k}\cdot\bs{q}}\; K_j(\bs{k},\bs{q})=P^{\text{E}}_{j}(\bs{k}),
\end{split}
\end{equation} 
where $P^{\text{E}}_{j}(k)$ is the $j$-order contribution to the Eulerian power spectrum (with the appropriate EFT corrections). 

This expression can be further simplified be introducing the Fourier transform of the Eulerian power spectrum. After doing so, the resummed correlation function is simply given by the convolution integral
\begin{equation}\label{eq.masterfinal}
\begin{split}
\xi(r)=\sum_{j=1}^N\int \dx{3}{q}\; \xi_j^\text{E}(q) \,R_{N-j}(\bs{r}-\bs{q},\bs{r}),
\end{split}
\end{equation}
where we have defined
\begin{equation}\label{eq.R}
\begin{split}
R_{N-j}(\bs{y},\bs{r})=\int \frac{\dx{3}{k}}{(2\pi)^3} \,e^{i \bs{k}\cdot\bs{y}}\left.\left.F(\bs{k},\bs{r})\right|\right|_{N-j}.
\end{split}
\end{equation}
The Fourier transform in Eq.~(\ref{eq.R}) can be done analytically and the explicit expression can be found in Appendix~\ref{app.angularintegral}. Furthermore, also the angular integral appearing in Eqs.~(\ref{eq.masterapprox}) and (\ref{eq.masterfinal}) can be calculated analytically, leaving only a one dimensional integral to be evaluated numerically. For $N=1$, that is IR-resummation performed on the tree-level two-point function, the expression takes the particularly simple form
\begin{equation}\label{eq.N=1}
\begin{split}
\xi_{\text{tree}}(r)=\frac{1}{|A_{ij}(\bs{r})|^{1/2}}\int\dx{3}{q}\;\xi_{\text{tree}}^\text{E}(q) \,\exp\[-\frac{1}{2}(\bs{r}-\bs{q})_iA_{ij}^{-1}(\bs{r})(\bs{r}-\bs{q})_j\].
\end{split}
\end{equation}

In the expressions in Eqs.~(\ref{eq.masterapprox}) and (\ref{eq.masterfinal}), $F$ is a $\bs{r}$-dependent Guassian-like kernel that modifies the Fourier transform of the Eulerian power spectrum, smoothing out the BAO wiggles. The fact that one obtains a good agreement with N-body simulations by suppressing the wiggly part of the power spectrum is not a novelty. However we would like to stress that the derivation proposed in \cite{Senatore:2014via} and simplified (and improved with subleading corrections)  here, does not require an arbitrary splitting in wiggly and non-wiggly power spectrum. It is therefore systematic, and provides control of theoretical uncertainty.  

Let us make one further remark about the expression in Eq.~(\ref{eq.masterapprox}). As one can see, the integrand on the right hand side of this expression coincides to the Eulerian power spectrum up to order $N$ in $P_{\text{lin}}$ but has additional terms proportional to the displacement at higher order; in other words, the only difference between the resummed expression and the perturbative Eulerian expression starts at an oder that is higher than the order of the Eulerian expression we use. The higher order terms account for the non-linear effect of long displacements (whose perturbative corrections are also automatically encoded in the formula).\\

We are now in a position to  include the additional NLO terms\footnote{Notice that the form of subleading terms is not universal (i.e. dictated by the Equivalence Principle \cite{Baldauf:2015xfa}), but has to be explicitly calculated from the equation of motion. See App.~\ref{app.AB} for the explicit form.} in the resumming kernel $K_0$. As discussed in Eq.~(\ref{eq.pertdispl}) (and below it) the most-IR-enhanced NLO terms are contained in the tree-level cubic cumulant of the long displacement field. Therefore the improved $K_0$ kernel reads
\begin{equation}\label{eq.K0W}
\begin{split}
K_0(\bs{k},\bs{q})\equiv \left.K(\bs{k},\bs{q})\right|_0&= \exp\[\sum^3_{n=1}\frac{(-i)^n}{n!}\<(\bs{k}\cdot\bs{\Delta^{\text{IR}}}(\bs{q}))^n\>_c\]\\&
=\exp\[-\frac{1}{2}k_ik_j A_{ij}^{\text{IR}}(\bs{q})+\frac{i}{6}k_ik_j k_k B_{ijk}^{\text{IR}}(\bs{q})\].
\end{split}
\end{equation}
The derivation just proposed follow straightforwardly since the same approximation of Eq.~(\ref{eq.s(x)}) in the third order cumulant is again of order $\epsilon_{\delta_<}$. In Sec.~\ref{sec.result} we will show quantitatively that these terms modify the IR-resummation only marginally, as expected from the parametric scaling in Eq.~(\ref{eq.NLOoverLoopShort}). \\

To close this section let us also write the power spectrum by exploiting the new resummation just proposed. By directly Fourier transforming the resummed correlation function of Eq.~(\ref{eq.masterapprox}) one obtains the resummed power spectrum. By adding and subtracting the Eulerian correlation function we can then write 
\begin{equation}\label{eq.deltapower}
\begin{split}
P(k)=\int \rd^3 r \,e^{i\bs{k}\cdot \bs{r}} \xi(r) = P^\text{E}(k) +\int \rd^3 r \,e^{i\bs{k}\cdot \bs{r}} (\xi(r)-\xi^E(r)).
\end{split}
\end{equation}
This expression has some advantages with respect to the usual IR resummation of  \cite{Senatore:2014via}. In that scheme, one is sensitive to a delta function in $k$-space which in real space amounts to having an integrand with support up to very large distances. Instead,  our expression in Eq.~(\ref{eq.deltapower}) is not sensitive to large distances $r$. Quite on the contrary, the expression has support only around the BAO peak, making the entire procedure quite economical and extremely easy to evaluate numerically with an 1D FFTLog \cite{Hamilton:1999uv}. In a sense, the splitting in Eq.~(\ref{eq.deltapower}) is similar to the wiggle/no-wiggle splitting often used in the literature. In fact, focussing on the wiggly part amounts to focusing in real space to the BAO peak. However, unlike such splitting, the manipulation in Eq.~(\ref{eq.deltapower}) is exact, and does not introduce unrecoverable systematic error in the resummation procedure.

\section{IR-resummation in redshift space}

In this section, following \cite{Senatore:2014vja, Lewandowski:2015ziq}, we derive the redshift space IR-resummation for the correlation function using the novel simplification proposed in this paper. 
Starting again from the Lagrangian picture, the overdensity in redshift space $\delta^r$ can be written as \cite{Matsubara:2007wj}
\begin{equation}\label{eq.deltaredshift}
\begin{split}
1+\delta^r(\bs{x}_r,t)=\int\rd^3q\,\delta_D^3(\bs{x}^r-\bs{q}-\bs{s}^r(\bs{q},t)),
\end{split}
\end{equation}
where in the planar approximation the displacement in redshift space $\bs{s}_r$ is given by
\begin{equation}\label{eq.dispredshift}
\begin{split}
\bs{s}^r=\bs{s}+ \frac{\bs{\hat{z}}\cdot\bs{\dot{s}}}{H}\bs{\hat{z}}.
\end{split}
\end{equation}
In this section we will be interested in resumming only the IR-modes of the linear displacement field, for which $\bs{\dot{s}}_{\text{lin}}\simeq f H \bs{s}_{\text{lin}}$, with $f=\partial \log D/\partial \log a$ and $D$ the growth factor. Using the linear solution, Eq.~(\ref{eq.dispredshift}) becomes
\begin{equation}\label{eq.lindispredshift}
\begin{split}
\bs{s}^r=\bs{s}+ f\bs{\hat{z}}\cdot\bs{s}\,\bs{\hat{z}}=\mathcal{R}\cdot \bs{s},
\end{split}
\end{equation}
where
\begin{equation}\label{eq.Red}
\begin{split}
\mathcal{R}_{ij}=\delta_{ij}+ f \hat{z}_j \hat{z}_j.
\end{split}
\end{equation}
This last factor modifies the two-point function of the long-displacement modes entering in the resummation kernel $K_0$ of Eq.~(\ref{eq.K0}), leading to the redshift-space resumming kernel $K_0^r$ \cite{Lewandowski:2015ziq},
\begin{equation}\label{eq.K0redshift}
\begin{split}
K_0^r(\bs{k},\bs{q})\equiv \left.K^r(\bs{k},\bs{q})\right|_0&=\exp\[-\frac{1}{2}k_ik_j \tilde{A}_{ij}^{\text{IR}}(\bs{q})\],
\end{split}
\end{equation}
where the correlation function of long displacement fields $A$ of Eq.~(\ref{eq.ABdecomposition}) is replaced by $\tilde{A}$ as follows
\begin{equation}\label{eq.Atilde}
\begin{split}
\tilde{A}= \mathcal{R}\cdot A\cdot\mathcal{R}.
\end{split}
\end{equation}
Going trough the same steps as before, we can directly write the correlation function in redshift space as
\begin{equation}\label{eq.masterfinalRedshift}
\begin{split}
\xi^r(r,\mu_r)=\sum_{j=1}^N\int \rd^3 q\, \xi_{j}^{\text{E},r}(q,\mu_q) \,\tilde{R}_{N-j}(\bs{r}-\bs{q},\bs{r}),
\end{split}
\end{equation}
where $\mu_v=\bs{\hat{v}} \cdot \bs{\hat{z}}$,  $\xi_{j}^{\text{E},r}$ is the $j$-th order of the correlation function in Eulerian PT in redshift space and $\tilde{R}$ is given by Eq.~(\ref{eq.R}) after the replacement $A\rightarrow\tilde{A}$. Again, this provides a simplification with respect to the formulas of \cite{Lewandowski:2015ziq}, as we discussed for the real space case.

At this point we can expand both correlations function in the multipole basis 
\begin{equation}\label{eq.?}
\begin{split}
\xi^r(r,\mu_r)=\sum_\ell\xi^r_\ell(r) \mathcal{P}_\ell(\mu_r),
\end{split}
\end{equation}
and write the multipoles of the correlation function as 
\begin{equation}\label{eq.masterfinalRedshift}
\begin{split}
\xi^r_\ell(r)=\sum_{j=1}^N\int \rd q\, q^2\, \xi_{j,\ell'}^{\text{E},r}(q) \,\tilde{R}_{N-j}^{\ell\ell'}(r,q),
\end{split}
\end{equation}
where
\begin{equation}\label{eq.Rll'}
\begin{split}
\tilde{R}_{N-j}^{\ell\ell'}(r,q)=\frac{2\ell+1}{2}\int_{-1}^{1} \rd \mu_r\int\rd^2 \Omega_q\,\tilde{R}_{N-j}(\bs{r}-\bs{q},\bs{r})  \mathcal{P}_\ell(\mu_r)  \mathcal{P}_{\ell'}(\mu_q).
\end{split}
\end{equation}
This 1+2-dimensional angular integral can be evaluated numerically for every $\ell$, $\ell'$, similarly to what is done for the redshift-space Zel'dovich approximation. 

\section{Results and Conclusions}\label{sec.result}
We now turn to the quantitative evaluation of the IR-resummation derived in Sec.~\ref{sec.lagrangian}. The leading effect, broadening and shifting the BAO peak, is due to modes of the displacement field of wavelengths smaller than $\ell_\text{BAO}$, and it was already captured by the resummation of \cite{Senatore:2014via}. Notice however that here we have simplified the original derivation, making the numerical evaluation of the IR-resummation  easier. Additionally, we also include the NLO dynamical effects due to short/long mode-coupling to confirm that in our Universe these terms have a small effect. We will compare the IR-resummation at zero-, one- and two-loop in the counting of Eulerian PT, but, as already stated, at each order the kernel $K_0$ of Eq.~(\ref{eq.K0}) resums all orders in the long displacement. Going to higher loops in the Eulerian PT correlation function has therefore two effects: first, it includes the higher order terms in $\epsilon_{\delta_<}$ and $\epsilon_{s_>}$ (which also corrects the description of the long displacement modes); second, it diminishes the residual explicit dependence on $\Lambda_\text{IR}$.

\begin{figure}[!htb]
  \centering
  \subfloat[]{\includegraphics[width=0.49\textwidth]{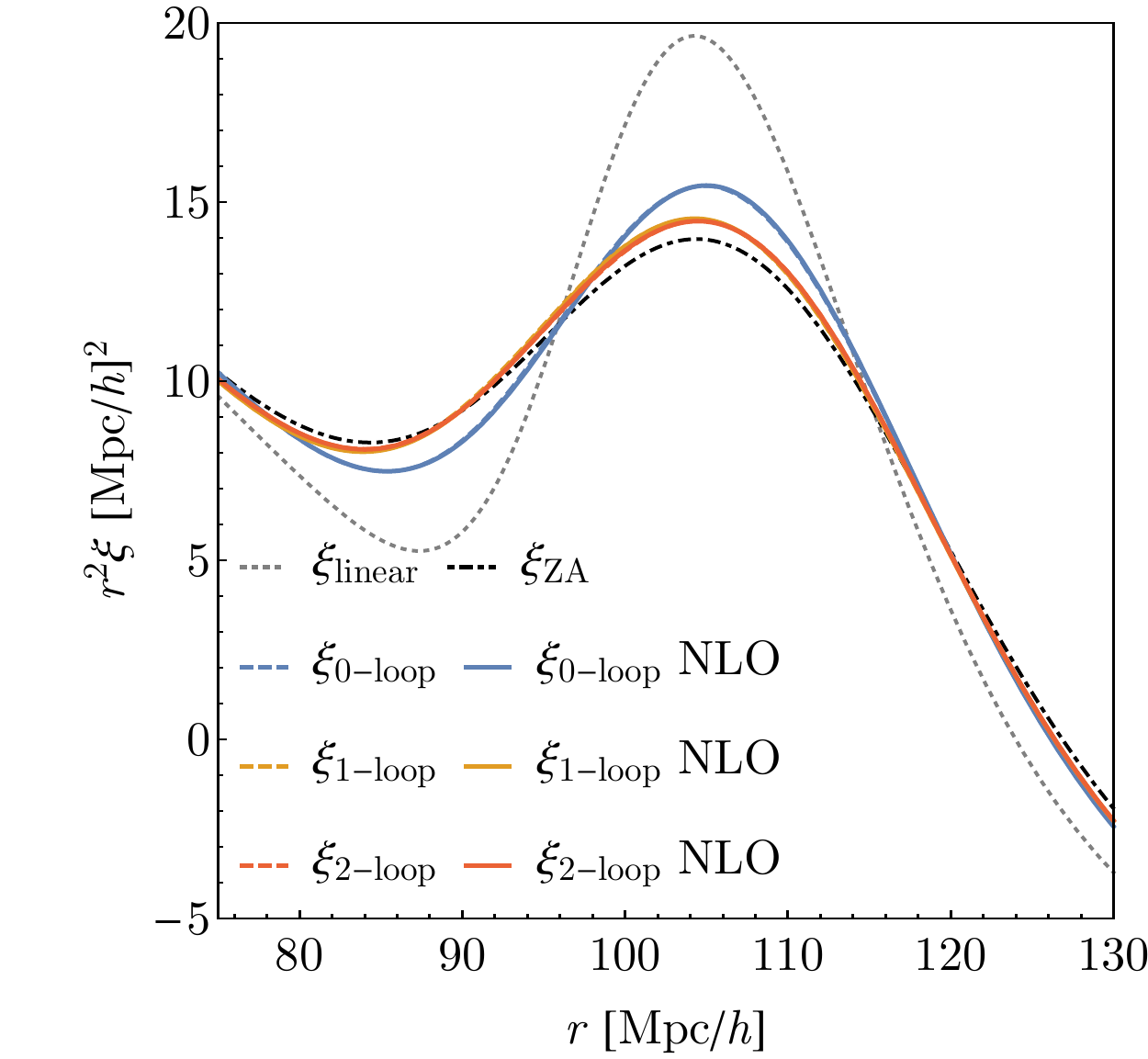}\label{fig.CorrFunc}}
  \hfill
  \subfloat[]{\includegraphics[width=0.49\textwidth]{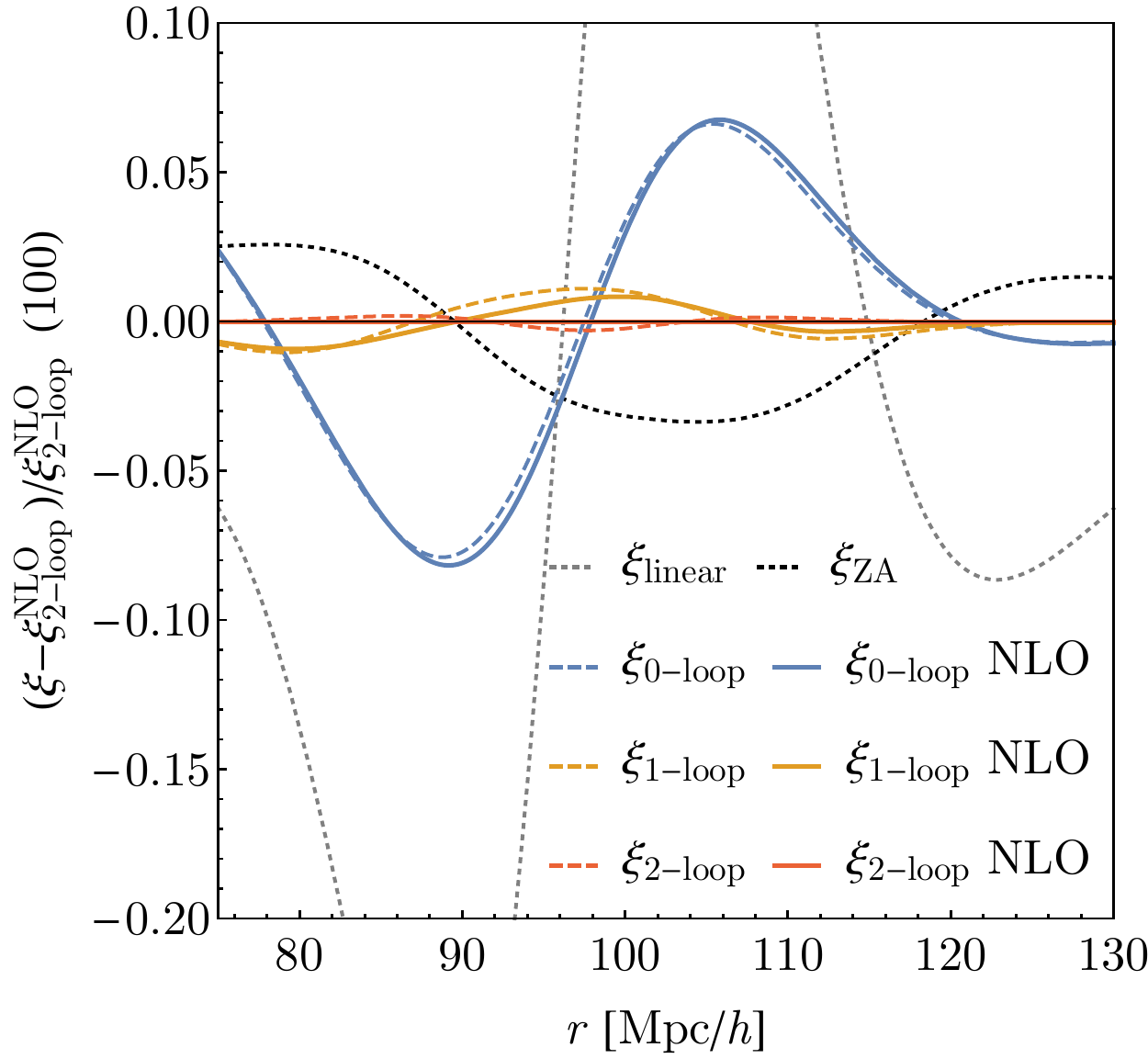}\label{fig.RelDiff}}
\caption{\small 
(a) {\em IR-resummed correlation function at zero-, one-, two-loop (respectively blue, orange, red) with and without the cubic terms discussed in the text (solid and dashed).}
(b) {\em Relative difference normalized at $r=100 \,h^{-1}\text{Mpc}$. In both cases the cutoff is $\Lambda_{\text{IR}}=0.12\,h\,\text{Mpc}^{-1}$. We see that the effect of the inclusion of the NLO is not larger than the inclusion of the higher order terms in $\delta$, as expected from the estimates done in the text. Furthermore, the convergence of the perturbative treatment is quite rapid. The code that produces these plots is available on the \href{http://stanford.edu/~senatore/}{EFTofLSS repository}.}}
  \label{fig.corr}
  \end{figure}

As we have shown in Subsec.~\ref{sec.main}, our  derivation of the IR-resummed PT calculates the correlation function directly in real space, so let us start with this. To evaluate the IR-resummation we use the linear power spectrum of the \href{http://darksky.slac.stanford.edu}{Dark Sky} simulation\footnote{The Dark Sky simulation (\href{http://darksky.slac.stanford.edu}{http://darksky.slac.stanford.edu}) evolved $10240^3$ particles in a volume of $(8\,h^{-1}\text{Gpc})^3$ with cosmological parameters  $\{\Omega_m,\Omega_b,\Omega_\Lambda,h, n_s, \sigma_8\} = \{0.295, 0.0468, 0.705, 0.688, 0.9676, 0.835\}$.} \cite{Skillman:2014qca}, and use the results of~\cite{Foreman:2015lca} which fitted the EFT parameters to the same simulation. Since we are not aware of measurements of the real-space correlation function for this simulation, in real space we will just study the convergence of the resummed correlation function as a function of the number of loops, also including the additional NLO terms defined at the end of Subsec.~\ref{sec.main} in the resummation kernel. 
In Fig.~\ref{fig.CorrFunc} and Fig.~\ref{fig.RelDiff} we compare the IR resummation in real space with and without the resummation of the cubic cumulant. As in~\cite{Senatore:2014via}, the convergence of the IR-resummed correlation function is around 10\% at tree-level and around 1\% at one-loop (compared to the best calculation we have, the two-loop + NLO terms), allowing us to estimate that the two-loop calculation is extremely close to the non-linear result. As we have already stated there is some residual freedom in choosing the IR cut-off for the IR-resummation; however as the number of loops increases, this residual freedom is reabsorbed. In principle one can use this freedom to adjust the IR cut-off in such a way to make the convergence faster, since the error will anyway cancel by going to higher order in loops. We choose $\Lambda_{\text{IR}}=0.12\,h\,\text{Mpc}^{-1}$ in order to match the one-loop result to the two-loop correlation function at $r=105\,h^{-1}\text{Mpc}$. With this choice the one-loop calculation is well within 1\% with respect to the two-loop calculation. 

\begin{figure}[!htb]
\begin{center}
\includegraphics[width=0.485 \textwidth]{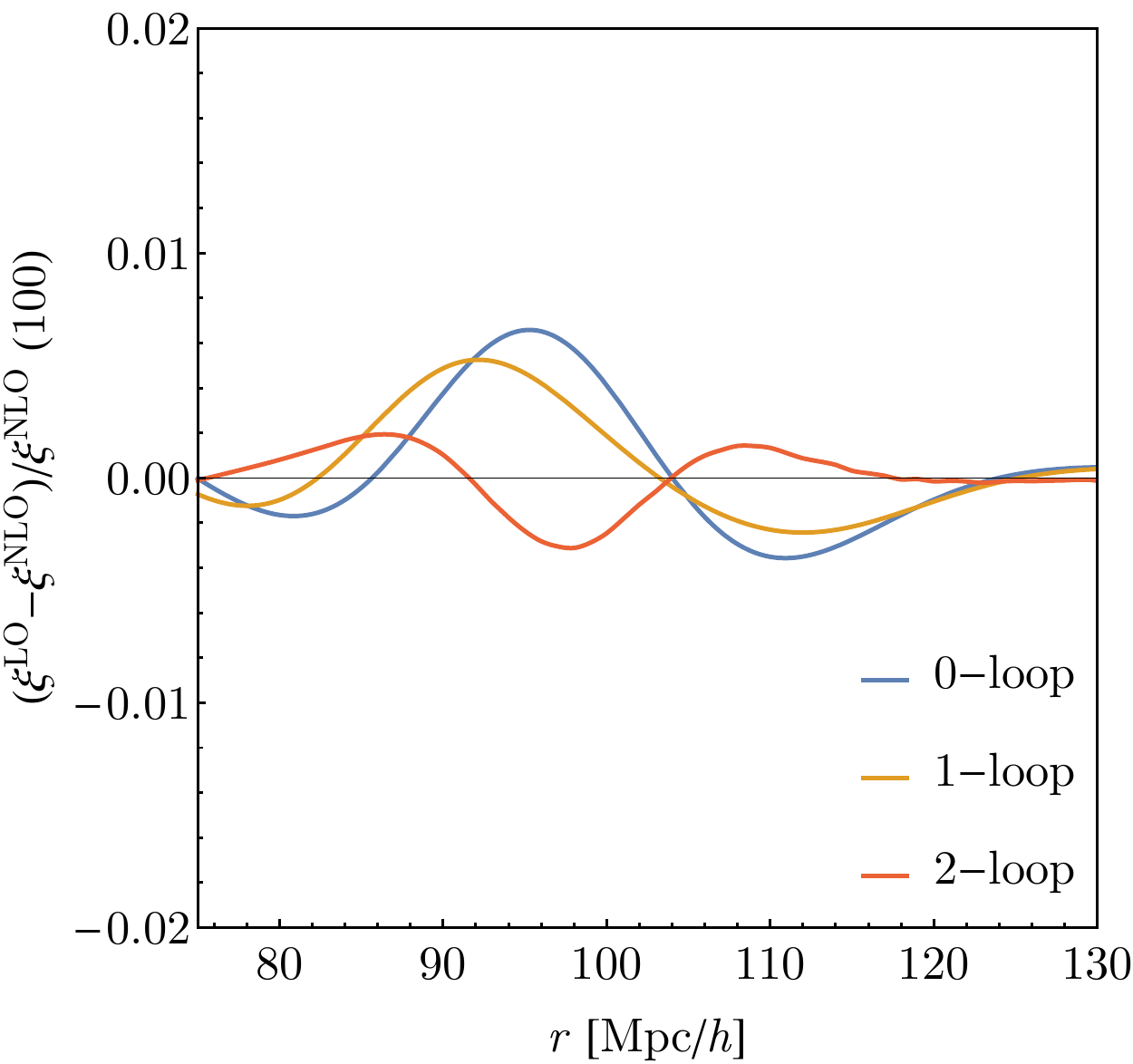}
\end{center}
\caption{\small {\em Relative change in the resummation when the cubic term is included, using a mix setting of infrared cutoffs: $\Lambda_{\text{IR}}=0.1\,h\,\text{Mpc}^{-1}$ for the LO term, $\Lambda_{\text{IR}}=0.5\,h\,\text{Mpc}^{-1}$ for the NLO term. The code that produces this plot is available on the \href{http://stanford.edu/~senatore/}{EFTofLSS repository}.}}
\label{fig.relchange}
\end{figure}

As it can be seen in Figs.~\ref{fig.corr} and \ref{fig.relchange}, the additional NLO terms, although parametrically IR-enhanced, do not modify the correlation function for more than 1\% (only $\sim0.25$\% at two-loop), improving only slightly the convergence of the one-loop calculation towards the two-loop. As anticipated, this is primarily due to the smallness of the cubic terms, which make the expansion of the exponential quickly convergent, and the suppression with respect to insertions of $\epsilon_{\delta_{<,\text{high}}}$, as in Eq.~(\ref{eq.NLOoverLoopShort}). This shows that, although one cannot speed up the convergence of PT by including these terms, the shape of the BAO feature is a robust prediction of the IR-resummed correlation function. 

\begin{figure}[!htb]
\begin{center}
\includegraphics[width=1 \textwidth]{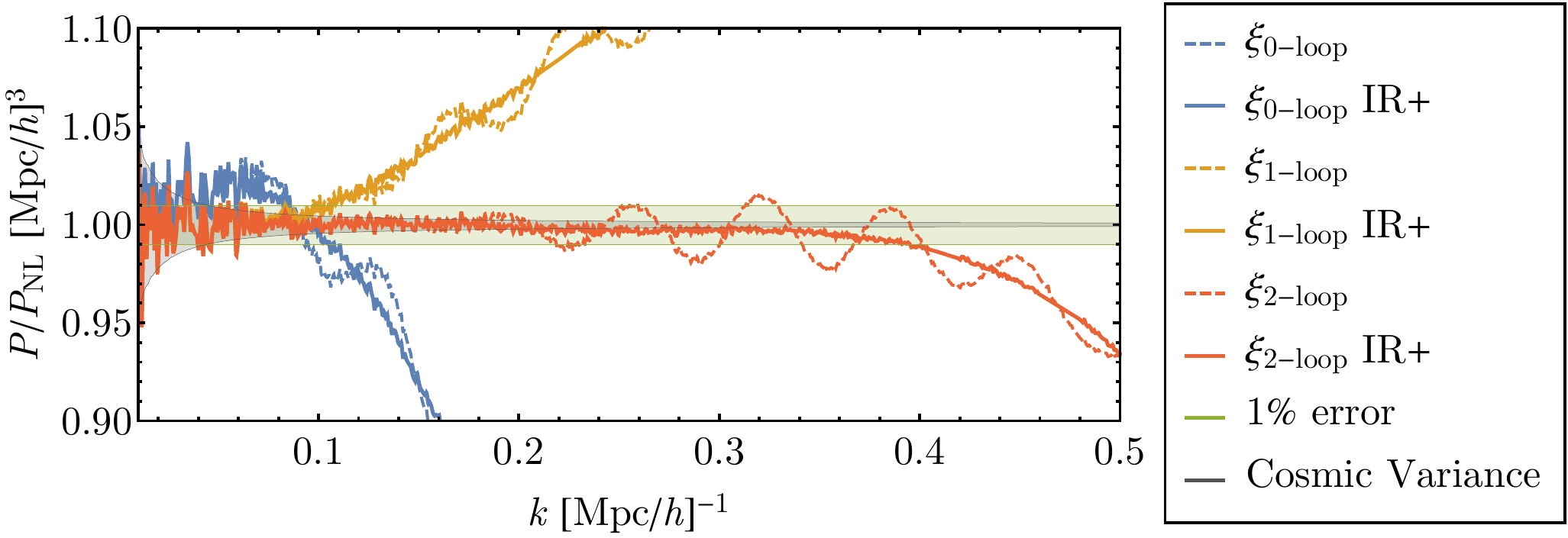}
\end{center}
\caption{\small {\em Power spectra at zero-, one- and two-loop, before and after the IR-resummation with $\Lambda_{\text{IR}}=0.12\,h^{-1}\text{Mpc}$. The IR-resummation improves the prediction for the wiggles in the power spectrum, leaving only an error in the broad band power. Data are taken from the }\texttt{ds14\_a} {\em run of \href{http://darksky.slac.stanford.edu}{Dark Sky}.}}
\label{fig.power}
\end{figure}

Let us now turn to Fourier space. In Fig.~\ref{fig.power} we show the result of the procedure proposed here against the power spectrum measured from the Dark Sky simulation\footnote{We used the power spectrum at zero redshift measured from the \texttt{ds14\_a} run of \href{http://darksky.slac.stanford.edu}{Dark Sky}. }.  Notice that here, contrary to what plotted in \cite{Foreman:2015lca}, we do not cancel the leading order cosmic variance (represented by the gray band in Fig.~\ref{fig.power}) since we are mostly interested in the wiggly component of the power spectrum which has support where cosmic variance is already  quite  small. 
As it can be clearly seen, the IR-resummation completely eliminates any residual oscillation from the ratio with respect to the measured power spectrum indicating that the theoretical power spectrum predicts very accurately the wiggly component due the baryonic oscillations (in real space this fact translates to the absence of the spurious peaks of Fig.~\ref{fig.PTBAO}). Notice that the IR-resummation works even at $k$ where perturbation theory starts to fail. This however should be of no surprise since, as we argued, we are non-perturbatively taking care of an infrared effect. Notice also that the IR-resummation does not (and should not) improved the overall reach of the theoretical prediction which is dominated by UV-physics (and not IR-physics).\\

Although we have not evaluated the IR-resummation proposed here in redshift space, we believe that similar benefits in the numerical evaluation of the resummation should be achieved also in redshift space, but leave this check for future works. 

\pagebreak
\section*{Acknowledgements}
We would like to thank Simon Foreman for providing the EFT power spectra. We would also like to thank Mehrdad Mirbabayi, Marko Simonovi\'c and Duccio Pappadopulo for useful discussions. G. T. is supported by the James Arthur Postdoctoral Fellowship. L. S. is partially supported by NSF award 1720397. 
\pagebreak
\appendix
\section{Correlation functions of the displacement field}\label{app.AB}
The functions entering in Eq.~(\ref{eq.ABdecomposition}) can be expressed by simple one-dimensional integrals as \cite{Matsubara:2007wj}
\begin{equation}\label{eq.AB}
\begin{split}
A_0(q)&=\frac{2}{3}(\Xi_0(q)-\Xi_2(q)),\\
A_2(q)&=2\,\Xi_2(q)),\\
B_1(q)&=\frac{1}{5}(2\,\Xi_1(q)-3\,\Xi_3(q)),\\
B_3(q)&=3\,\Xi_3(q)),\\
\end{split}
\end{equation}
where, at tree-level,
\begin{equation}\label{eq.xi}
\begin{split}
\Xi_0(q)&=\int\frac{\rd p}{2\pi^2}\, P_{\text{lin}}(p) (1-j_0(pq)),\\
\Xi_1(q)&=\int\frac{\rd p}{2\pi^2}\,\(-\frac{3}{7p}\) \(Q_1(p)-3Q_2(p)+2R_1(p)-6R_2(p)\)j_1(pq),\\
\Xi_2(q)&=\int\frac{\rd p}{2\pi^2}\,P_{\text{lin}}(p)\, j_2(pq),\\
\Xi_3(q)&=\int\frac{\rd p}{2\pi^2}\,\(-\frac{3}{7p}\)\(Q_1(p)+2Q_2(p)+2R_1(p)+4R_2(p)\)j_3(pq),\\
\end{split}
\end{equation}
and
\begin{equation}\label{eq.?}
\begin{split}
Q_n(p) &=  \frac{p^3}{4\pi^2}\int_0^\infty \rd r\,P_{\rm lin}(pr) \int_{-1}^1 \rd\mu P_{\rm lin}[k(1 + r^2 - 2r\mu )^{1/2}] \frac{\tilde{Q}_n(r,p)}{(1 + r^2 - 2r\mu)^2},\\
R_n(p) &= \frac{p^3}{4\pi^2}P_{\rm lin}(p) 
    \int_0^\infty \rd r\, P_{\rm lin}(pr)\,\tilde{R}_n(r),
\end{split}
\end{equation}
where
\begin{equation}\label{eq.?}
\begin{split}
\tilde{Q}_1 &= r^2 (1 - x^2)^2,\\
\tilde{Q}_2 &= (1 - x^2) rx (1 - rx),\\
\tilde{R}_1 &= -\frac{1}{24r^2}(1 + r^2)(3 - 14 r^2 + 3 r^4)+ \frac{3}{24r^3} (r^2 - 1)^4 \ln\left|\frac{1+r}{1-r}\right|,\\
 \tilde{R}_2 &=\frac{1}{24r^2}(1 - r^2)(3 - 2 r^2 + 3 r^4)+ \frac{3}{24r^3} (r^2 - 1)^3 (1 + r^2) \ln\left|\frac{1+r}{1-r}\right|.
\end{split}
\end{equation}

\section{Explicit form for the resumming kernel $R$}\label{app.angularintegral}
The evaluation of the correlation function in Eq.~(\ref{eq.masterfinal}) involves the the following quantity 
\begin{equation}\label{eq.Mrepeat}
\begin{split}
R_{N-j}(\bs{y},\bs{r})&=\int \frac{\rd^3 k}{(2\pi)^3} \,e^{i \bs{k}\cdot\bs{y}}\left.\left.F(\bs{k},\bs{r})\right|\right|_{N-j}\\
&=\int \frac{\rd^3 k}{(2\pi)^3} \,e^{i \bs{k}\cdot\bs{y}}K_0(\bs{k},\bs{r})\left.\left.K_0^{-1}(\bs{k},\bs{r})\right|\right|_{N-j},
\end{split}
\end{equation}
where $K_0$ is nearly Gaussian. When the kernel $K_0$ is gaussian (in $\bs{k}$), which is the case for the leading order resummation (see Eq.~(\ref{eq.K0})), one can compute exactly the Fourier transform which reads
\begin{equation}\label{eq.Mexplicit}
\begin{split}
R_{N-j}(\bs{y},\bs{r})&=\left.\left.K_0^{-1}(-i\nabla_{\bs{y}},\bs{r})\right|\right|_{N-j}G(\bs{y},\bs{r}),
\end{split}
\end{equation}  
where
\begin{equation}\label{eq.gauss}
\begin{split}
G(\bs{y},\bs{r})&=\frac{1}{|A_{ij}(\bs{r})|^{1/2}}\exp\[-\frac{1}{2}y_iy_j A_{ij}^{-1}(\bs{r})\].
\end{split}
\end{equation}
At this point the angular integral in Eq.~(\ref{eq.masterfinal}) can be easily done analytically, leaving us with just a one-dimensional numerical integral. \\

When the cubic term is added (see Eq.~(\ref{eq.K0W})) one can compute the integral in Eq.~(\ref{eq.Mrepeat}) by  expanding the cubic term appearing in the exponent and evaluating the same Gaussian integrals of Eq.~(\ref{eq.Mexplicit}). The result is 
\begin{equation}\label{eq.Mexplicit}
\begin{split}
R_{N-j}(\bs{y},\bs{r})&=\sum_{m=0}^{\infty}\frac{1}{m!}\[C(-i\nabla_{\bs{y}},\bs{r})\]^m\left.\left.K_0^{-1}(-i\nabla_{\bs{y}},\bs{r})\right|\right|_{N-j}G(\bs{y},\bs{r}),
\end{split}
\end{equation}  
where
\begin{equation}\label{eq.cubicterm}
\begin{split}
C(\bs{k},\bs{r})&=\frac{i}{6}k_ik_j k_k B_{ijk}^{\text{IR}}(\bs{r}).
\end{split}
\end{equation}
Since the dependence on the cubic is analytic and the overall magnitude of $C$ is much smaller than one, the series in Eq.~(\ref{eq.Mexplicit}) converges  rapidly.

\section{Alternative formulas for the power spectrum}\label{app.M}
In this Appendix we write an alternative but equivalent formula for the power spectrum. Let us also write the power spectrum by Fourier transforming Eq.~(\ref{eq.masterapprox})
\begin{equation}\label{eq.masterPower}
\begin{split}
P(k)
&=\sum_{j=1}^N\int \frac{\rd^3 k'}{(2\pi)^3}\,P^{\text{E}}_{j}(\bs{k'})\int \rd^3 r \,e^{i (\bs{k'}-\bs{k})\cdot\bs{r}}\left.\left.F\(\bs{k'},\bs{r}\)\right|\right|_{N-j},\\
&=\sum_{j=1}^N\int \frac{\rd^3 k'}{(2\pi)^3}\,P^{\text{E}}_{j}(\bs{k'})M_{N-j}(\bs{k},\bs{k'}),\\
\end{split}
\end{equation}
where 
\begin{equation}\label{eq.M}
\begin{split}
M_{N-j}(\bs{k},\bs{k'})=\int \rd^3 r \,e^{i (\bs{k'}-\bs{k})\cdot\bs{r}}\, \left.\left.F\(\bs{k'},\bs{r}\)\right|\right|_{N-j}.
\end{split}
\end{equation}

Had we not performed the IR-resummation, the matrix $M$ would simply be a Dirac delta-function of $\bs{k}-\bs{k'}$, leaving us with the power spectrum of the Eulerian EFT. However the IR-resummation adds to the Eulerian calculation higher order terms which improve the prediction for the wiggles in the power spectrum.\\

 Eq.~(\ref{eq.masterPower}) can be further simplified as
\begin{equation}\label{eq.masterPowerinAPP}
\begin{split}
P(k)
&=\sum_{j=1}^N\int \frac{\rd^3 k'}{(2\pi)^3}\,P^{\text{E}}_{j}(\bs{k'})M_{N-j}(\bs{k},\bs{k'})\\
&=\sum_{j=1}^N\int \rd k' \,k'^2P^{\text{E}}_{j}(k') \bar{M}_{N-j}(k,k'),\\
\end{split}
\end{equation}
where
\begin{equation}\label{eq.MinApp}
\begin{split}
\bar{M}_{N-j}(k,k')&=\frac{1}{{(2\pi)^3}}\int\rd \Omega_{{\bs{k}'}}M_{N-j}(\bs{k},\bs{k'})\\
&=\frac{1}{{(2\pi)^3}}\int\rd^3 r\int \rd \Omega_{{\bs{k}'}} \,e^{i (\bs{k'}-\bs{k})\cdot\bs{r}}\, F\(\bs{k'},\bs{r}\)||_{N-j}\\
&=\frac{1}{{(2\pi)^3}}\int\rd^3 r \,e^{-i\bs{k}\cdot\bs{r}}\int \rd \Omega_{{\bs{k}'}} e^{i \bs{k'}\cdot\bs{r}}\left.\left.F\(\bs{k'},\bs{r}\)\right|\right|_{N-j}\\
&=\frac{1}{{(2\pi)^3}}\int\rd^3 r \,e^{-i\bs{k}\cdot\bs{r}} \bar F\(k',r\)||_{N-j}\\
&=\frac{1}{{2\pi^2}}\int\rd r \,r^2 \,j_0(kr)\bar F\(k',r\)||_{N-j},\\
\end{split}
\end{equation}
where 
\begin{equation}\label{eq.Fbar}
\begin{split}
 \bar F\(k',r\)||_{N-j}&=\int \rd \Omega_{{\bs{k}'}} e^{i \bs{k'}\cdot\bs{r}}\left.\left.F\(\bs{k'},\bs{r}\)\right|\right|_{N-j}.\\
\end{split}
\end{equation}
Notice that attention should be paid in evaluating numerically the above integral since at large $r$, $K_0(k,r)\rightarrow K_0(k,\infty)= e^{-k^2 A_0(\infty)/2}$, so that the integral in Eq.~(\ref{eq.MinApp}) contains a delta function. One trick to evaluate the power spectrum consists in rewriting Eq.~(\ref{eq.masterPowerinAPP}) by adding and subtracting the contribution at infinity as
\begin{equation}\label{eq.newP}
\begin{split}
P(k)
&=\sum_{j=1}^N\int \frac{\rd^3 k'}{(2\pi)^3}\,P^{\text{E}}_{j}(\bs{k'})(M_{N-j}(\bs{k},\bs{k'})- M_{N-j}^{\infty}(\bs{k},\bs{k'}))\\
&+\sum_{j=1}^N\int \frac{\rd^3 k'}{(2\pi)^3}\,P^{\text{E}}_{j}(\bs{k'})M_{N-j}^{\infty}(\bs{k},\bs{k'}),\\
\end{split}
\end{equation}
where 
\begin{equation}\label{eq.Minfinity}
\begin{split}
M_{N-j}^{\infty}(\bs{k},\bs{k'})&=\int \rd^3 r \,e^{i (\bs{k'}-\bs{k})\cdot\bs{r}}\, \left.\left.F\(k',\infty\)\right|\right|_{N-j},\\
 \left.\left.F\(k',\infty\)\right|\right|_{N-j}&= e^{-k'^2 A_0^{\text{IR}}(\infty)/2}\cdot e^{k'^2 A_0^{\text{IR}}(\infty)/2}||_{N-j}.
\end{split}
\end{equation}
With this Eq.~(\ref{eq.newP}) becomes
\begin{equation}\label{eq.?}
\begin{split}
P(k)
&=\sum_{j=1}^N\int \frac{\rd^3 k'}{(2\pi)^3}\,P^{\text{E}}_{j}(\bs{k'})(M_{N-j}(\bs{k},\bs{k'})- M_{N-j}^{\infty}(\bs{k},\bs{k'}))\\
&+\sum_{j=1}^N e^{-k^2 A_0^{\text{IR}}(\infty)/2}\cdot e^{k^2 A_0^{\text{IR}}(\infty)/2}||_{N-j}P^{\text{E}}_{j}(k),\\
\end{split}
\end{equation}
where now the integrand in the first line falls off rapidly at $r\rightarrow\infty$ and can be easily evaluated numerically. \\

\bibliographystyle{JHEP}
\bibliography{mybib}
\end{document}